# Resonant Nonlinear Damping of Quantized Spin Waves in Ferromagnetic Nanowires: A Spin-Torque Ferromagnetic Resonance Study

C. T. Boone[1], J. A. Katine[2], J. R. Childress[2], V. Tiberkevich[3], A. Slavin[3], J. Zhu[1], X. Cheng[1],

I. N. Krivorotov[1]

1. *Department of Physics and Astronomy, University of California, Irvine, California 92697*
2. *Hitachi Global Storage Technologies, San Jose, California 95135*
3. *Department of Physics, Oakland University, Rochester, Michigan 48309*

## Abstract

**We use spin torque ferromagnetic resonance to measure the spectral properties of dipole-exchange spin waves in permalloy nanowires. Our measurements reveal that geometric confinement has a profound effect on the damping of spin waves in the nanowire geometry. The damping parameter of the lowest-energy quantized spin wave mode depends on applied magnetic field in a resonant way and exhibits a maximum at a field that increases with decreasing nanowire width. This enhancement of damping originates from a nonlinear resonant three-magnon confluence process allowed at a particular bias field value determined by quantization of the spin wave spectrum in the nanowire geometry.**

Spin transfer torque (STT) [1-4] is emerging as a new tool for studies of magnetization dynamics in nanostructures [5-13]. In this Letter, we use STT to measure properties of dipole-exchange spin waves in permalloy (Py = $Ni_{86}Fe_{14}$) nanowires. Our measurements reveal that geometric confinement of spin waves in nanowires has a profound effect on the spin wave damping [14]. We find a resonant dependence of the damping parameter of the lowest-energy spin wave mode on external magnetic field − a maximum of damping is observed at a field that increases with decreasing nanowire width. We explain this resonant enhancement of damping by a nonlinear three-magnon confluence process in which two quanta of the first ($n$=1) spin wave mode merge into a single quantum of the third ($n$ =3) mode.

Brillouin light scattering [15,16], ferromagnetic resonance (FMR) [17-20] and magneto-optical techniques [21,22] have been applied to studies of spin waves in ferromagnetic wires. Here we use spin torque FMR (st-FMR) [5,6], to make measurements of the spectral properties of dipole-exchange spin waves in Py nanowires. In this technique, a ferromagnetic nanocontact is used to apply an ac spin-polarized current, $I_{ac}$, to a metallic nanowire on a nonmagnetic metallic substrate as illustrated in Fig. 1. When the frequency of the current flowing perpendicular to the nanowire into the substrate is equal to the eigen-frequency of a spin wave mode, this mode is resonantly excited by the ac STT from the current. The excited mode induces resistance oscillations at the nanocontact due to the giant magnetoresistance effect and generates a rectified voltage, $V_{dc}$, proportional to the mode amplitude [5,6]. As the frequency of the ac current, $f$, is swept through a spin wave eigen-frequency, a resonance peak or a trough appears in the plot of $V_{dc}$ versus $f$. The peaks and troughs in the st-FRM spectrum, $V_{dc}(f)$, give information about spin wave frequencies, amplitudes and damping parameters.

We make st-FMR measurements for a set of eight 6-nm thick and 100-250 nm wide Py nanowires lithographically defined on a Cu/Ta underlayer that serves as the bottom electrical lead. Spin-polarized current is injected into the nanowire through a $Co_{50}Fe_{50}$/ Ru/ $Co_{50}Fe_{50}$/ $Ir_{20}Cr_3Mn_{77}$ synthetic antiferromagnet (SAF) nanopillar patterned on top of the Py nanowire and separated from the nanowire by an 8-nm thick Cu spacer, as shown in Fig. 1. The top Au/Ta lead is connected to the SAF fixed layer and current is applied between the top and the bottom leads.

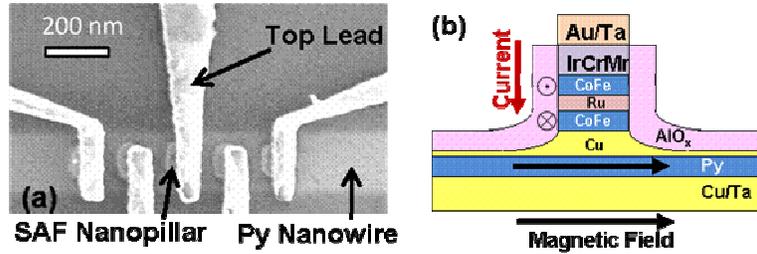

Fig. 1. (color online): (a) Scanning electron microscopy (SEM) image of a Py nanowire device with five current-polarizing SAF nanocontacts and Au/Ta top leads. (b) Schematic side view of the sample.

Since the Cu spacer layer above the Py layer is only partially milled away to protect the Py nanowire surface, some lateral current spreading in the spacer takes place. Five spin current injector nanocontacts are attached to the Py nanowire. In this Letter, we report single-contact st-FMR measurements. The device in Fig. 1(a) is made in a multi-step nanofabrication process starting from a Ta(5)/ Cu(30)/ Ta(3)/ Cu(30)/ Ta(5)/ Cu(3)/ Py(6)/ Cu(8)/ $Co_{50}Fe_{50}$ (3.5)/ Ru(0.8)/ $Co_{50}Fe_{50}$ (3.5)/ $Ir_{20}Cr_3Mn_{77}$ (7)/ Cu(10)/ Ru(5)/ Ta(2.5) multilayer (thicknesses are in nanometers). The values of saturation magnetization of the Py ($M = 580$ emu/cm$^3$) and $Co_{50}Fe_{50}$ (1480 emu/cm$^3$) films are measured by vibrating sample magnetometry. The nanowire axis is perpendicular to the SAF exchange bias direction in order to maximize the magnitude of STT [6]. For st-FMR measurements, we apply modulated microwave current to the nanocontact and measure the resulting rectified voltage, $V_{dc}$, using lock-in detection [6]. The frequency of the current is swept in the 1–15 GHz range and, after subtraction of a background voltage due to Ohmic heating [13], the resulting $V_{dc}(f)$ gives the st-FMR spectrum. Fig. 2(a) shows a series of st-FMR spectra measured at several fields applied along the nanowire axis of a 240 nm wide Py nanowire at T = 4.2 K. These spectra exhibit three resonances (peaks and troughs) corresponding to the lowest three spin wave modes in the nanowire, as shown in Fig. 2(b). All three modes shift to higher frequencies, and the frequency spacing among the modes decreases with increasing magnetic field [15]. The observed resonances are due to spin waves in the Py nanowire since the lowest-frequency SAF acoustic mode at H = 0 is expected at $f > 13$ GHz − well above the frequencies of the resonances in Fig. 2(b) [23]. The line shape of each mode can be fit to a sum of a symmetric and an antisymmetric Lorentzians, indicating that not only STT but also magnetic field from the top lead contribute to excitation of spin waves [13]. From the amplitude $V_{dc}(f_1)$ of the lowest-energy mode we estimate the precession cone angle of this mode to be 4.6 °/mA. In our sample geometry, spin waves with wave vectors ranging from 0 to $k_x^{max}$ can be excited, where $k_x$ is the wave vector along the nanowire axis, and the value of $k_x^{max}$ is determined by the extent of current spreading in the Cu spacer. Therefore, the line shape of each mode in the st-FMR spectra is a convolution of line shapes of spin waves with $k_x < k_x^{max}$. For our samples this inhomogeneous line broadening does not exceed line broadening due to intrinsic spin wave damping as discussed later in this Letter. This allows us to identify positions of minima and maxima of the resonances in st-FMR spectra with the frequencies of spin wave modes at $k_x = 0$.

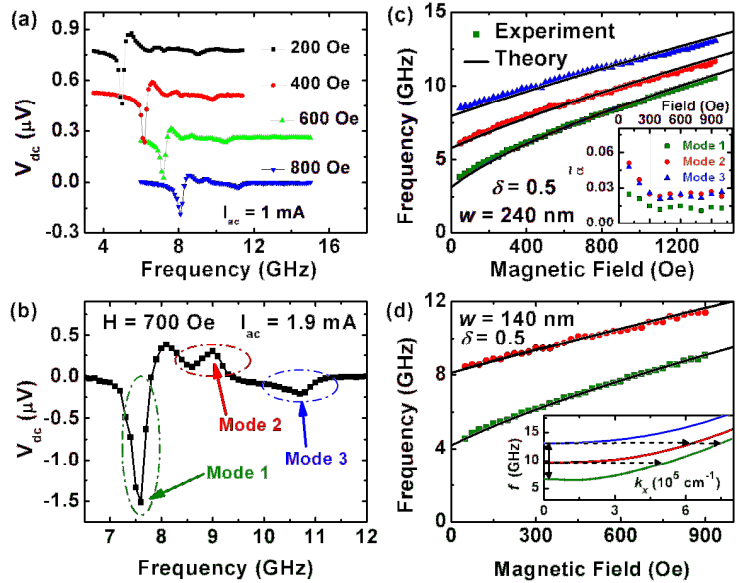

Fig. 2. (color online): (a) Evolution of st-FMR spectra with increasing magnetic field applied parallel to a 240 nm wide nanowire. The curves are vertically offset for clarity. (b) st-FMR spectrum showing three spin wave resonances. (c), (d): Mode frequency versus field applied parallel to the nanowire for a 140 nm wide wire and a 240 nm wide wire. Symbols are data and lines are Eq. (1) fits with the nanowire width $w$ and the pinning parameter $\delta$ as fitting parameters. Inset in (c) is the dependence of the damping parameters of three spin wave modes in the 240 nm wide nanowire on magnetic field. Inset in (d) shows the dispersion relation calculated from Eq. (1) for a 140 nm wire at 400 Oe. The dashed arrows indicate the two-magnon scattering channels and the solid arrow shows the $k_x=0$ three-magnon confluence process active only at the field for which the frequency of the n=1 mode is half the frequency of the n=3 mode.

To quantify the effect of confinement on spin waves in nanowires, we compare the measured spin wave frequencies to a theoretical expression for the frequency of quantized dipole-exchange spin wave width modes in thin-film strips of rectangular cross section [24,25]:

$$f_n = \frac{\gamma}{2\pi} 4\pi M \sqrt{\left[\frac{H}{4\pi M} + \frac{A}{2\pi M^2}\kappa_n^2 + \left(1 - \frac{1-\exp(-\kappa_n d)}{\kappa_n d}\right)\frac{k_{yn}^2}{\kappa_n^2}\right]\left[\frac{H}{4\pi M} + \frac{A}{2\pi M^2}\kappa_n^2 + \frac{1-\exp(-\kappa_n d)}{\kappa_n d}\right]} \quad (1)$$

In this equation, $f_n$ is the frequency of $n$-th mode, $\gamma/2\pi = 2.95$ MHz/Oe is the gyromagnetic ratio, $d$ is the nanowire thickness, $H$ is the magnetic field along the strip axis, $A = 10^{-6}$ erg/cm is the exchange constant, and $\kappa_n^2 = k_x^2 + k_{yn}^2$ where $k_{yn}$ is the quantized spin-wave wave vector along the width of the strip. In our measurements, both even and odd modes are excited due to inhomogeneous current injection over the nanowire width arising from the current injector being misaligned with the nanowire center as shown in Fig. 1(a). The quantized wave vector, $k_{yn}$, generally can be expressed as $k_{yn} = \frac{(n-\delta_n)\pi}{w}$, where $w$ is the nanowire width and $0 \leq \delta_n \leq 1$ is a pinning parameter determined by the boundary conditions for dynamic magnetization at the nanowire edges ($\delta_n = 0$ corresponds to complete pinning, whereas $\delta_n = 1$ describes free spins). We fit Eq. (1) to the $f_n(H)$ data using $\delta_n$ and $w$ as fitting parameters, assuming that $\delta_n$ is independent of $n$ and is identical for all studied nanowires, $\delta_n = \delta$. We use $w$ as a fitting parameter because we find ~ 20% spread in the spin wave frequencies for nanowires of nominally identical width. This fitting procedure gives high-quality fits shown in Fig. 2(c) and 2(d) for $f_n(H)$ with $\delta \approx 0.5$ and the nanowire width $w$ similar to the nanowire width measured with SEM (135 nm and 225 nm with ~ 10% standard deviation for two groups of the nanowires studied). A recent theory of boundary conditions for dipole-exchange spin waves in thin-film elements [25] predicts values of $\delta$ smaller ($\delta \approx 0.2$, corresponding to stronger edge pinning) than $\delta = 0.5$ found in our experiment. We expect some edge pinning reduction for nanowires with thicknesses comparable to or smaller than the exchange length of Py (~ 5 nm) (see Fig. 2 in [25]). The edge pinning in our system can be also reduced because the magnetic properties of Py near the nanowire edges are modified by ion milling [26, 27], and, therefore, the assumptions of spatially homogeneous magnetization and exchange used in [25] are not satisfied for our system.

Fitting a mode spectral line shape to a linear combination of symmetric and antisymmetric Lorentzians [13] gives the mode half width at half maximum, $\Delta f_n$, which carries information about the mode damping parameter, $\alpha_n$. Due to inhomogeneous line broadening caused by excitation of spin waves with $k_x$ in the $(0 - k_x^{max})$ range, $\Delta f_n$ sets an upper bound on $\alpha_n$:
$$\alpha_n \leq \frac{2\pi \Delta f_n}{\gamma(H + 2\pi M)} \equiv \tilde{\alpha}_n \text{ [28]}.$$

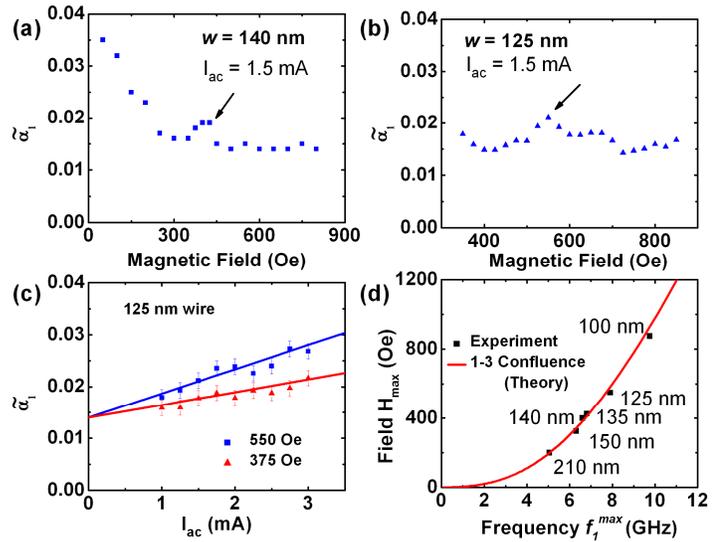

Fig. 3 (color online): (a), (b) Damping parameter of the first mode $\tilde{\alpha}_1$ versus field for 140 nm and 125 nm nanowires, showing a maximum of damping at a field $H_{max}$ (425 Oe and 550 Oe, respectively). (c) Damping versus ac drive current for the 125 nm wire at $H = H_{max}$ (550 Oe) and $H < H_{max}$ (375 Oe). (d) $H_{max}$ versus frequency of the first mode at $H = H_{max}$. Squares are data, line shows the field at which frequency of the third mode is equal to twice the frequency of the first mode as calculated from Eq. (1), data labels show the nanowire width obtained from Eq. (1) fit of the $f_n(H)$ data.

The inset in Fig. 2(c) shows $\tilde{\alpha}_n$ for the first three modes of a nanowire with $w$ = 240 nm as a function of field. The high-field ($H$ > 250 Oe) value of $\tilde{\alpha}_1$ =0.013 ± 0.002 does not significantly exceed the value of Gilbert damping parameter $\alpha \approx$ 0.01 for thin Py films [28]. Using the dispersion relation of Eq. (1) and assuming intrinsic damping $\alpha$ =0.008 [12], inhomogeneous line broadening gives $\tilde{\alpha}_1$ =0.013 for $k_x^{max}$ = 2×10$^5$ cm$^{-1}$. The width of the region where spin torque is applied to the nanowire, $\pi/k_x^{max}$=160 nm, is approximately twice as large as the physical width of the current injector, which can be explained by current spreading in the Cu spacer. We observe a large increase of $\tilde{\alpha}_n$ for $H$ < 250 Oe for all modes. A low-field increase of damping of similar magnitude was previously observed in thin Py films [28,29]. This damping enhancement can be attributed to enhanced inhomogeneous broadening [30] and to incomplete saturation of magnetization at low fields [27].

The inset in Fig. 2(c) also demonstrates that the damping parameters of the $n \geq 2$ modes are enhanced compared to damping of the $n$=1 mode. This increase of damping can be explained by two-magnon scattering of spin waves [31,32]. Higher order ($n \geq 2$) width modes can scatter into lower order modes with the same frequency and $k_x \neq$0 via a two-magnon process leading to the linewidth broadening. A likely mechanism for the observed large two-magnon scattering is a random magnetic potential induced by the nanowire edge roughness or by magnetization pinning centers. In contrast, two-magnon scattering of the $n$=1 mode to other modes is prohibited by conservation of energy, and the high-field value of the damping parameter for this mode is close to the Py thin-film value.

For nanowires with $w$ < 220 nm, we observe a non-monotonic dependence of damping of the $n$=1 mode on magnetic field as shown in Fig. 3(a) and 3(b). The damping as a function of field exhibits a maximum at a field $H_{max}$ that increases with decreasing $w$. This damping maximum is a nonlinear effect as demonstrated in Fig. 3(c) that shows $\tilde{\alpha}_1$ versus $I_{ac}$ measured at $H_{max}$=550 Oe and at $H = 375$ Oe $< H_{max}$ for a 125-nm wide nanowire. The damping increases linearly with $I_{ac}$ (and thus with the mode amplitude) for all field values, but the rate of increase is significantly larger at $H = H_{max}$, indicating that an additional nonlinear damping-enhancing process is active at $H = H_{max}$. Fig. 3(d) shows the dependence of $H_{max}$ on frequency of the first mode at this field, $f_1^{max}$, for nanowires of different widths. We also note that at $H_{max}$, the frequency of the lowest-energy mode, $f_1$, is equal to half the frequency of the third mode, $2f_1 = f_3$. The solid line in Fig. 3(d) shows the field at which $2f_1 = f_3$ as a function of $f_1$ calculated from Eq. (1). This theoretical curve is in excellent agreement with the observed dependence of $H_{max}$ on $f_1^{max}$. Therefore, we conclude that the observed enhancement of damping at $H_{max}$ is due to a three-magnon confluence process in which two spin waves of the $n=1$ mode merge into a single spin wave of the $n=3$ mode [33,34]. Due to quantization of the spin wave spectrum in nanowires, this energy- and momentum-conserving scattering channel has a resonant character and exists only at a particular magnetic field, $H_{max}$.

In summary, we use st-FMR to study spin waves in Py nanowires and find that two-magnon scattering enhances the damping of higher order quantized spin wave width modes compared to the damping of the lowest-energy mode. We also observe that the damping of the lowest-energy width mode resonantly depends on the bias magnetic field and exhibits a maximum at a particular field value. We attribute this maximum to a nonlinear three-magnon confluence process in which two magnons of the first width mode merge into a single magnon of the third width mode. This resonant dependence of the spin wave damping parameter on magnetic field originates from quantization of the spin wave spectrum in the nanowire geometry.

Our work demonstrates that the nonlinear three-magnon confluence process creates an additional resonant damping channel for the lowest-energy spin wave mode of a ferromagnetic nanowire and gives rise to an unusual dependence of spin wave damping on external magnetic field. This work was supported by the NSF, and by the NRI through the Western Institute of Nanoelectronics.